\documentclass[11pt,a4paper, preprint,  superscriptaddress]{revtex4}
\usepackage[utf8]{inputenc}
\usepackage{hyperref}
\usepackage{pdflscape}
\usepackage{romannum}
\usepackage{caption}
\usepackage{amsmath}
\usepackage{amssymb}
\usepackage{subfigure}
\usepackage{epsfig}
\usepackage{epstopdf} 
\usepackage{amsmath,amsfonts}
\usepackage{graphicx}
\begin{document}
	\title{\Large   A simplistic approach to the study of two-point correlation function in galaxy clusters}
	\author {Durakhshan Ashraf Qadri}
	\email{durakhshanqadrii@gmail.com}
	\affiliation{{Department of Physics, National Institute of Technology  Srinagar, Jammu and Kashmir -190006, India.}}
	\author {Abdul W. Khanday}
	\email{abdulwakeelkhanday@gmail.com}
	\affiliation{{Department of Physics, National Institute of Technology  Srinagar, Jammu and Kashmir -190006, India.}}
	
	\author {Prince A. Ganai}  
	\email{princeganai@nitsri.net}
	\affiliation{{Department of Physics, National Institute of Technology  Srinagar, 
			Jammu and Kashmir -190006, India.}}

	\begin{abstract}
		We developed the functional form of the two-point correlation function under the approximation of fixed particle number density $\bar{n}$.We solved the quasi-linear partial differential equation (PDE) through the method of characteristics to obtain the parametric solution for the canonical ensemble. We attempted many functional forms and concluded that the functional form should be such that the two-point  correlation function should go to zero as the value of system temperature increases or the separation between the galaxies becomes large . Also we studied the graphical behavior of the developed two-point correlation function for large values of temperature $T$ and spatial separation $r$. The behavior of the two-point function was also studied from the temperature measurement of clusters in the red-shift range of $0.023 - 0.546$.
	\end{abstract}	
	\maketitle
	\section{Introduction}
		Galaxies in clusters serve as robust cosmological observatories and special astrophysical laboratories. Thus, they provide a veritable understanding about the Universe at large scale. The Universe exhibits the hierarchical behavior which exists at all scales, from the smallest quantum particles to the ultimately vast structures, with galaxy clusters occupying the top pyramid of the structure formation. Being the virialized structures in the universe,clusters of galaxies are the largest gravitationally bound objects in the universe.
	 The distribution of the matter at large scales is dominated by the gravitational interaction and the gravitational clustering of galaxies plays an important role in the evolution of the Universe. The studies of gravitational clustering have the advantage that the rules do not change as the system evolves thus making them the useful cosmological probes. A linear theory of the development of the structure from the initial isothermal and adiabatic density perturbations to the present day observed cosmic web has been developed extensively~\cite{Lifshitz}.  The current observations of Galaxy clusters indicate that their motions have been strongly influenced by their mutual gravitational dynamics \cite{Klypin} . Physical processes requiring a long and complicated sequence of events are responsible for the non-linear clustering phenomenon\cite{saslaw}. Because Clustering is a many-body gravitational problem involving billions of stars and thousands of galaxies, it cannot be solved in the same way as the standard two-body problem, so a statistical treatment is required, following an analogy between thermodynamics and statistical physics~\cite{iqbal}.  The concept of distribution function is fundamental to a statistical description of a dynamical system . The distribution and correlation functions define the overall clustering of galaxies. The implications of these ideas are also consistent with observations. The statistical mechanical study of the structure formation and distribution in the universe has been studied rigorously in ~\cite{khanday1,khanday2,khanday3, khanday4, upadhyay, hameeda}. The authors have employed various modified gravity theories to study the effects of these modifications on the statistical properties of large scale structures in the universe. \\
	 One of the most common approaches to study the origin of the Universe is to analyze correlation functions\cite{totsuji}. The correlation function is a general method for studying the distribution of galaxies in a cluster and is a ubiquitous tool for measuring the degree of clustering.\\
	The two-point correlation function is defined through the conditional probability of  finding a galaxy in some region $dV_2$ at distance $r$ from a second galaxy in region defined by volume $dV_1$, as\cite{peebles}
	\begin{equation}
	P_{2|1} = \bar{n}(1+\xi_2(r))dV_2,
	\end{equation}
	where $r=r_2-r_1$ is the spatial separation between the galaxies and $\xi_2$ is the two-point correlation function. In general, the two-point correlation function will depend on the absolute positions of the two galaxies i.e. $r_1$ and $r_2$. However, if we average over all directions  then $\xi_2$ becomes a function of $r=r_2-r_1$ for a statistically homogeneous system.
	The  nature of gravitational  interaction being pairwise , it directly depends on lower order correlation functions such as two-point correlation rather than higher order correlations.

	The development of functional form of the two-point correlation function has been attempted in \cite{naseer}. The authors of \cite{naseer} have developed the two-point correlation function with a variable number of the system particles. Similarly, for a single component system the the two-point correlation function has also been developed in \cite{farooq}.  Although this is a valid attempt to develop the function, yet the system is such that the appreciable change in the particle number takes place on a time scale larger than the relaxation time of the cluster of galaxies. \\
	The development of the two-point correlation function keeping the particle number fixed is the aim of the present work. We focus on the variation of the correlation function with a changing system temperature $T$ and inter-particle separation $r$ keeping the particle number $N$ fixed.     
		
	 This paper has been arranged as follows. In section (\Romannum{2}) we develop a differential equation of  the two-point correlation function  keeping the number density $\bar{n}$ constant. In section (\Romannum{3}) we propose the functional form of the differential equation developed in section (\Romannum{2}) and we also define the uniqueness of the functional form and therefore the behavior of clusters in the expanding universe. In section (\Romannum{4}) we study the graphical behavior of the function for varying system temperature $T$ and spatial separation $r$. In section (\Romannum{5}) we study the two-point function using data analysis. Finally, in section (\Romannum{6}) we make the  discussion and conclusion.

	\section {Development of differential equation for two-point correlation function}
	
	We assume an infinite system of point galaxies having same mass $m$ in order to keep the system uniform. For such a system of assumed point particles interacting gravitationally, the internal energy , $U$ and pressure, $P$  satisfy the standard statistical thermodynamic relations;
	\begin{equation}
		U=\frac{3}{2}NT(1-b)\label{1} 
	\end{equation}
	\begin{equation}
   	P = \frac{NT}{V}(1-2b)\label{2},
    \end{equation}
	
	where $N$ represents the average number of particles in a canonical ensemble given by;	
\begin{equation}
N ={\bar{n}}V \notag
\end{equation}
and 
\begin{equation}
		\bar{n} = \frac{N}{V},\notag
\end{equation}

	where $\bar{n}$ represents the average number density of the  system particles.
	
	Equations ( \ref{1}) and (\ref{2}) are the equations of State, and $b$ is a dimensionless parameter, a measure of the ratio of the gravitational correlation energy and the kinetic energy due to peculiar velocities$(K=\frac{3}{2}NT)$ and is given by;(\cite{farooq}).
	$$b=-\frac{W}{2K}$$

	As discussed above for $\xi_2$ to provide a good description, the system should be statistically homogeneous throughout, so that average for any sufficiently large subset of galaxies and that of a small subset in the same volume will be the same.
	
	For a grand canonical ensemble, in which particle number changes,  two-point correlation function $\xi_2$ depends on $\bar{n}$, $T$ and $r$ i.e., $	\xi_2 = \xi_2(\bar{n}, r, T)$
	and the variation in $\xi_2$ for such a system can be written as;
	\begin{equation}
	d\xi_2 =  \frac{\partial\xi_2}{\partial T } dT +  \frac{\partial\xi_2}{\partial r } dr+  \frac{\partial\xi_2}{\partial \bar{n} } d\bar{n} 
		\end{equation}

It is a well established fact that the number of  particles (galaxies) does not  change appreciably in a cluster unless a collision occurs. Hence, it is justified to fix the  particle number density $\bar{n}$. It can also be seen from the behavior of chemical potential in a cluster of galaxies, see e.g.,\cite{khanday2}. Thus we fix $\bar{n}$  such that;
	$$\frac{\partial\xi_2}{\partial \bar {n}}  = 0$$.
	Under this assumption the differential equation for the two-point correlation function $\xi_2(r, T)$ can be written as;
	\begin{equation}
 T \frac{\partial\xi_2}{\partial T} -r \frac{\partial\xi_2}{\partial r}=0\label{7}.
\end{equation}	
	Equation (\ref{7}) is the reduced first order partial differential equation for two-point correlation function.

\section{The possible functional form for the two-point correlation function $\xi_2$ }
	
	The equation (\ref{7}) is of the form of  Quasi-linear Partial differential equation. Hence method of characteristics is an important tool for solving hyperbolic-type partial differential equations (PDEs). Through this method, special curves known as the Characteristics Curves are determined along which the partial differential equation becomes the family of Ordinary differential equations (ODEs). Once the ODEs are obtained, they can be solved along the characteristics curves to get the solutions and thus can be related to the solution of the original PDEs.	
	While we attempt for the development of the functional form of $\xi_2$, the following observed boundary conditions must be satisfied;

	1.  In a homogeneous universe, the gravitational clustering of galaxies requires $\xi_2$ to have be positive  for some limiting values of  $\bar{n}$ , T and r.
	
	2.  For very low values of  $T$ and $r$  the correlation function  $\xi_2$ will increase for a constant value of $\bar{n}$ . Similarly for large value of T and r, the corresponding correlation function $\xi_2$ will decrease for constant value of $\bar{n}$ .
	
	3.  Due to virial equilibrium, galaxy clustering becomes more dominant as two-point correlation function $\xi_2$ increases, which implies that at low temperatures and high densities, more and more clusters are formed.\\
	
	4. The change in particle number becomes significant if a collision occurs, otherwise the particle number does not vary in a given volume.

 The system of equations defined by equation (\ref{7}) can be solved by the method of characteristics conveniently. We write our equation as;
\begin{equation}
 T \frac{\partial\xi_2}{\partial T} -r \frac{\partial\xi_2}{\partial r}=0\label{8}
 \end{equation}
The equation (\ref{8}) can further be written as;
\begin{equation}
 \frac{dT}{dr} = \frac{T}{-r}
\end{equation}
 After separation of variables and integrating, The equation (\ref{8}) leads to;

$$
\ln\left|T\right| = -\ln\left|r\right| + \ln\left|C1\right|$$
\begin{equation}
\implies C_1= T.r,\label{9}
\end{equation}
where $C_1$ is some unknown function and a solution to the PDE.\\
Again, we set  ;
$$\frac{\partial\xi_2}{\partial r} =0$$
\begin{equation}
\implies C_2=\xi_2,\label{10}
\end{equation}
where $c_2$ is another unknown function.

From equations (\ref{9}) and (\ref{10}) the functional form of $\xi_2$  has the following parametric structure;
\begin{equation}
	\xi_2(T,r) =f(T.r).\label{12}
\end{equation}
Equation (\ref{12}) will be used to obtain the exact functional form of the two-point correlation function. 


In order to explain the uniqueness of the devised functional form (\ref{12}), we test various combinations and finally choose the solution of equation(\ref{8}) as
\begin{equation}
 \xi_2(T,r) = \frac{\alpha_1}{1+\alpha_2 Tr},\label{11}
 \end{equation}
where $\alpha_1$ and $\alpha_2$ are free parameters that should be fixed through the comparison of this solution with the observational data.\\
The evolution of the two-point function with the red-shift is also important and  has been studied in\cite{koo}. The two-point correlation function satifies the following relation:
\begin{equation}
\xi_2(r,z)=\xi_2^0\left(1+z\right)^{-(3+\epsilon)}.\notag
\end{equation}

In a similar fashion we can also study the evolution of the correlation function by combining the well know relationship $r(t)= r_0(1+z)$ with equation \ref{11}, the two point correlation function \ref{11} can be written as
\begin{equation}
	\xi_2= \frac{\alpha_1}{1+\alpha_3T(1+z)},
\end{equation}
where ,$\alpha_3 = \alpha_2 r_0$, is again a parameter.

\section{Graphical representation of the two-point correlation function}
\begin{figure}[h!]
			\centering
			\includegraphics[width=8 cm, height=6 cm]{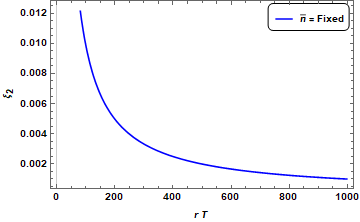}
			\captionof{figure}{Variation of the correlation function $\xi_2$ as a function of system temperature $T$ and mutual separation $r$ for fixed values of $\alpha_1$ and $\alpha_2$. }\label{fig1}
			
	\end{figure}

\begin{figure}[h!]
	\centering
	\includegraphics[width=8 cm, height=6 cm]{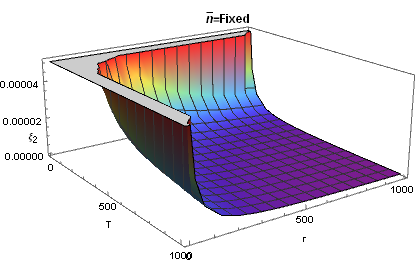}
	\caption{3D visualization of the behavior of correlation function $\xi_2$ as a function of system temperature $T$ and mutual separation $r$. }\label{fig2} 
\end{figure}

The behavior of the correlation function developed here can be visualized graphically from figure (\ref{fig1}). 
It is clear from the graph that as the value of the parameter $T.r$ increases the correlation decreases and tends to zero for relatively large values of $T.r$. This behavior is expected as we see that the galaxies far away from each other are less correlated and have minimum influence on one another. Similarly, the evolution of the correlation function with the gas temperature  and with an increasing separation is also shown in fig. (\ref{fig2}). It can be seen that the function shows a steep dip with increasing values of $T$ and $r$ independently.    
\section{Study of the Variation of correlation function with temperature through data }
\begin{figure}[h!]
	\centering
	\includegraphics[width=8 cm, height=6 cm]{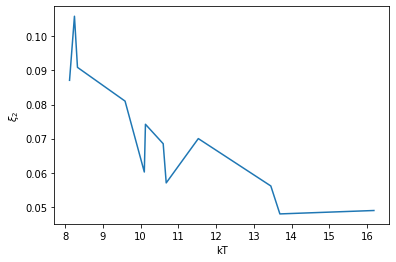}
	\caption{ Variation of the two-point correlation function with increasing temperature (kT(eV)).}\label{fig3}
\end{figure}

The variation of the two-point correlation function with system temperature shows an appreciable depreciation as can be seen from the graphical visualization, fig. (ref{fig1}). In terms of the  data measurements, the behavior of the correlation function with varying temperature of various clusters is also shown in the graph, fig.(\ref{fig3}). We have used the cluster temperature values from the work\cite{luiz}. The data includes measurements by Berkeley-Illinois-Maryland-Association(BIMA)and Owens Radio Observatory (OVRO). The values of cluster red-shifts, temperature and the corresponding values of the two-point correlation function is given in table (\ref{table}).  \\

\begin{figure}[h]
	\centering
	\subfigure[]{\includegraphics[width=0.48\textwidth]{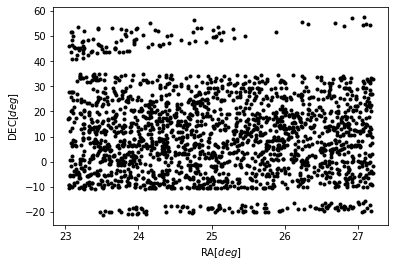}} 
	\subfigure[]{\includegraphics[width=0.48\textwidth]{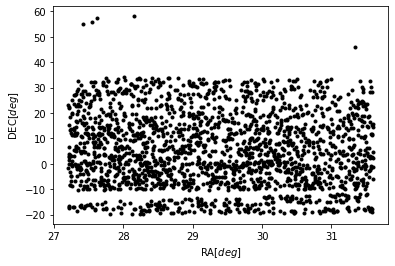}}\\
	\subfigure[]{\includegraphics[width=0.48\textwidth]{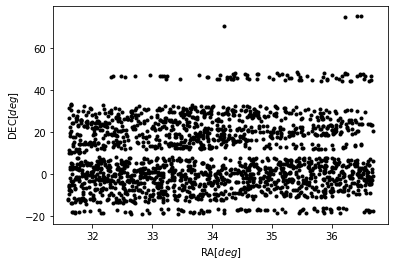}} 
	\subfigure[]{\includegraphics[width=0.48\textwidth]{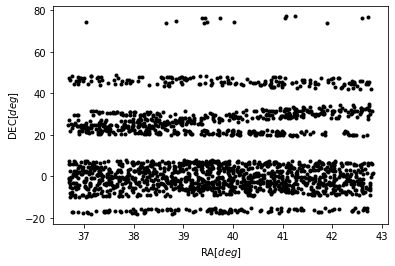}}\\ 
	\subfigure[]{\includegraphics[width=0.48\textwidth]{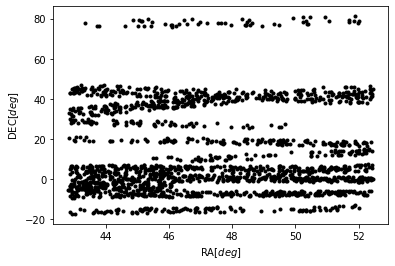}} 
	\subfigure[]{\includegraphics[width=0.48\textwidth]{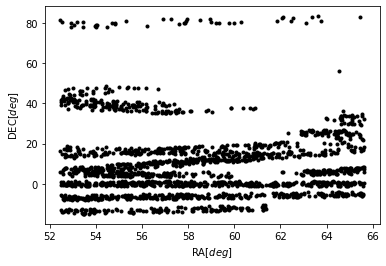}}
	
	\caption{(a-f) sky distribution in RA(deg) and DEC(deg) coordinates of galaxy clusters in various  red-shift ranges}
	\label{fig4}
\end{figure}

\begin{table}[ht]
\caption{ Cluster red-shifts(z), gas temperature(kT(eV)) and corresponding $\xi_2$ values }
	\begin{tabular}{|c| c| c| c|} 
		\hline
		Clusters & $z$ & $kT(eV)$ & $\xi_2$ \\ [0.5ex] 
		\hline
		A1656 & 0.023 & 6.62 & 0.1105 \\ 
		\hline
		A2204 & 0.152 & 8.12 & 0.0871 \\
		\hline
		A1689 & 0.183 & 8.25 & 0.1059 \\
		\hline
		A520 & 0.200 & 8.33 & 0.0909 \\  
		\hline
		A2163 & 0.202 & 9.59 & 0.081 \\ 
		\hline
		A773 & 0.216 & 10.1 & 0.0602 \\
		\hline
		A2390 & 0.232 & 10.13 & 0.0742 \\
		\hline
		A1835 & 0.252 & 10.6 & 0.0685 \\
		\hline
		A697 & 0.282 & 10.68 & 0.057 \\
		\hline
		ZW3146  & 0.291 & 11.53 & 0.07 \\
		\hline
		RXJ1347 & 0.451 & 13.45 & 0.0561 \\
		\hline
		CL0016+16 & 0.546 & 13.69 & 0.0479 \\
		\hline
		MS0451 - 0305 & 0.550 & 16.18 & 0.0489\\
		\hline
	\end{tabular}\label{table}
\end{table}
 From the graph we see that the correlation function decreases with increasing system temperature, fig. (\ref{fig3}). It is also observed that there are certain fluctuation at some red-shifts which may correspond to the clumpy distribution of galaxy clusters in some regions of space as seen in fig.(\ref{fig4}) .

\section{discussion and Conclusion}
In this paper we studied the two-point correlation function for a fixed particle number density $\bar{n}$. The approximation is valid because the change in the particle number in a cluster occurs on a very slow time scale. We treat our system in quasi-equilibrium and developed the functional form of the correlation function.After attempting many combinations , we choose the one in equation (\ref{11}). We also  visualized the graphical behavior of the correlation function for varying system temperature $T$ and spatial separation $r$. We also studied the variation of $\xi_2$ with a changing temperature for various clusters at different red-shifts, $0.023\leq z \leq 0.546$. From fig. (\ref{fig3}) we observed that the two-point function shows an appreciable decline with increasing system temperature that is predicted in equation \ref{11}. We also observed fluctuation at certain red-shifts whic can be possibly due to some unusual distribution in some regions of space as depicted in fig (\ref{fig4}).


\begin{thebibliography}{}
	

\bibitem{Lifshitz} Lifshitz, E. (2017). Republication of: On the gravitational stability of the expanding universe. General Relativity and Gravitation, 49(2), 1-20.
\bibitem{Klypin} Klypin, A. A., and Shandarin, S. F. (1983). Three-dimensional numerical model of the formation of large-scale structure in the Universe. Monthly Notices of the Royal Astronomical Society, 204(3), 891-907.
\bibitem{saslaw} Saslaw, W. C., and Hamilton, A. J. S. (1984). Thermodynamics and galaxy clustering-Nonlinear theory of high order correlations. The Astrophysical Journal, 276, 13-25.
\bibitem{iqbal}Iqbal, N., Ahmad, F., and Khan, M. S. (2006). Gravitational clustering of galaxies in an expanding universe. Journal of Astrophysics and Astronomy, 27(4), 373-379.
\bibitem{khanday1} Khanday, A. W., Upadhyay, S., and Ganai, P. A. (2021). Galactic clustering under power-law modified newtonian potential. General Relativity and Gravitation, 53(6), 1-19.
\bibitem{khanday2} Khanday, Abdul W.,   Upadhyay, S., and  Ganai, P. A. (2021). Thermodynamics of galaxy clusters in modified Newtonian potential. Physica Scripta 96, 125030.
\bibitem{khanday3} Khanday, Abdul W., Sudhaker Upadhyay, and Prince A. Ganai. "Statistical description of galactic clusters in Finzi gravity model." arXiv preprint arXiv:2203.17237 (2022).
\bibitem{khanday4} Khanday, Abdul W., Hilal A. Ganai, and Sudhaker Upadhyay. "Effect of nonfactorizable background geometry on thermodynamics of clustering of galaxies." arXiv preprint arXiv:2203.17243 (2022).
\bibitem{upadhyay} Upadhyay, S. (2017). Thermodynamics and galactic clustering with a modified gravitational potential. Physical Review D, 95(4), 043008.
\bibitem{hameeda} Hameeda, M., Pourhassan, B.,  Faizal, M.,   Masroor, C. P., Ansari,  R.-Ul H., and Suresh, P. K.(2019). Modified theory of gravity and clustering of multi-component system of galaxies. The European Physical Journal C   79, 769.
\bibitem{peebles}Peebles, Phillip James Edwin. The large-scale structure of the universe. Vol. 98. Princeton university press, 2020.
\bibitem{totsuji} Totsuji, Hiroo, and Taro Kihara. "The correlation function for the distribution of galaxies." Publications of the Astronomical Society of Japan 21 (1969): 221.
\bibitem{naseer}Iqbal, Naseer, Farooq Ahmad, and M. S. Khan. "Gravitational clustering of galaxies in an expanding universe." Journal of Astrophysics and Astronomy 27.4 (2006): 373-379.
\bibitem{farooq}Ahmad, Farooq. Two-particle correlation function and gravitational galaxy clustering. No. IUCAA-96-8. SCAN-9605099, 1996.
\bibitem{luiz}Luzzi, G., et al. "Redshift dependence of the cosmic Microwave Background temperature from Sunyaev-Zeldovich measurements." The Astrophysical Journal 705.2 (2009): 1122.
\bibitem{koo} Koo, D. C., and A. S. Szalay. "Angular correlations of galaxies to B= 24-Another probe of cosmology and galaxy evolution." The Astrophysical Journal 282 (1984): 390-397.
\end{thebibliography}
\end{document}